\documentstyle[12pt,amssymb,epsfig]{article}
\newcommand{\la}{\lambda}
\begin{document}
\begin{flushright}
KIAS Preprint P01023 \\
May 2001 \\
\end{flushright}
\vspace{1cm}
\begin{center}
{\Large\sc {\bf }}
\vspace*{3mm}
{\Large\sc {\bf Large mass splittings between charged and neutral
Higgs bosons in the MSSM}} 
\vspace{1cm}

{\large
{{A.G. Akeroyd}$^{\mbox{a,}}$\footnote{E--mail: akeroyd@kias.re.kr},
{S. Baek}$^{\mbox{b,}}$
\footnote{E--mail: swbaek@kias.re.kr}}}
\vspace{.6cm}
{\it
\\
a: Korea Institute for Advanced Study, 207-43 Cheongryangri-dong,\\
Dongdaemun-gu, Seoul 130-012, Korea\\

\vspace{.6cm}
b: Physics Department,
National Taiwan University,\\
10764 Taipei, Taiwan
}

\end{center}

\vspace{2cm}
\begin{abstract}
\noindent
We show that large ($\gtrsim 100$ GeV) mass splittings between the
charged Higgs boson ($H^\pm$) and the neutral Higgs bosons ($H^0$
and $A^0$) are possible in the Minimal Supersymmetric
Standard Model (MSSM).  Such splittings occur when the
$\mu$ parameter is considerably larger than 
the common SUSY scale, $M_{SUSY}$, and have significant consequences for 
MSSM Higgs searches at future colliders.

\end{abstract}

\newpage
\section{Introduction}
The Minimal Supersymmetric Standard Model (MSSM) \cite{prep} is currently
the leading candidate for physics beyond the Standard Model (SM). 
At the tree--level the Higgs sector
of the MSSM takes the form of a two Higgs doublet model (2HDM), 
where the coefficients of the 
quartic scalar terms are functions of the electroweak gauge 
couplings \cite{Gun}.
Five physical Higgs bosons are predicted: a charged pair ($H^+$,$H^-$),
two CP--even scalars ($h^0$,$H^0$), and a CP--odd $A^0$.
In the tree--level approximation the lightest Higgs $h^0$ satisfies 
$M_{h^0} \le M_Z$, while
$M_{H^\pm}\approx M_{H^0}\approx M_{A^0}$ for 
$M_{H^\pm}\gtrsim 200$ GeV.
Only two parameters are needed to fully parametrize the tree--level
potential, and these are usually taken to be $\tan\beta$ ($=v_2/v_1$,
where $v_i$ is the vacuum expectation value of a Higgs doublet) and
one of $M_{H^\pm},M_{A^0}$. 

At the 1--loop level the coefficients in the Higgs potential 
receive contributions from virtual particle loops and thus become
complicated functions of several (SUSY) parameters. 
In addition, new quartic scalar terms are generated.
This changes the tree--level mass relationships,
most notably weakening the above mass bound to $M_{h^0}\lesssim 130$ GeV
\cite{1-loop}. The near degeneracy relationship, 
$M_{H^\pm}\approx M_{A^0}\approx M_{H^0}$ for the large 
($\ge 200$ GeV) mass region, is only slightly affected, 
resulting in mass splittings of ${\cal O}(10)$ GeV. Recently there
has been much interest in the 1--loop effective potential of the MSSM with
unconstrained CP violating phases \cite{NPB553,PilWag,plb495,CPH}.
 It has been shown that
the mass splitting between the two heavier neutral scalars
(now mixed states of CP) can be increased up to 30 GeV for large phases.

In this paper we will show that large mass splittings 
between the charged and neutral Higgs bosons, 
$|M_{H^{\pm}}-M_{A^0,H^0}|\gtrsim 100$ GeV,
are possible in the CP conserving MSSM in a previously ignored
parameter space. These large splittings occur when 
$\mu$ is larger than the supersymmetric (SUSY) scale 
(defined as the arithmetic mean of the stop masses) 
by a factor of 4 or more. 
Although sizeable corrections to the tree--level mass 
sum rules can be implicitly found in previous works
e.g. \cite{Haber:1993an}, we identify the parameter space
for the largest splittings and discuss the phenomenological
consequences.
Importantly we show that this parameter space is consistent with
current experiments.

Such splittings would break the commonly assumed degeneracy relation 
$M_{H^\pm}\approx M_{H^0}\approx M_{A^0}$ for $M_{H^\pm}\gtrsim 200$ GeV, 
which should no longer be taken as a prediction of the MSSM.
The Higgs mass spectrum of the MSSM may resemble that of a general
2HDM or other extended (SUSY) Higgs sectors with scalar singlets etc. 
Knowledge of the maximum possible mass splittings among the 
Higgs bosons in the MSSM may be crucial
in distinguishing different models at future colliders.
In particular, measurements of the mass splittings provide important
information on the structure of the underlying Higgs sector, especially
since Higgs branching ratios (BR) and cross--sections 
are often very similar in many popular models. 
Moreover, new decays channels involving $H^\pm$,$H^0$ and $A^0$ would
be open and may possess large branching ratios, thus affecting 
proposed MSSM Higgs search strategies at future colliders.

Our work is organized as follows. In Section 2 we outline
our approach for evaluating the mass splittings.
Section 3 presents our numerical results, while section 4 
contains our conclusions.

\section{Mass splittings in the MSSM}


The most general CP violating Higgs potential of the MSSM may 
be described by the following effective Lagrangian:

\begin{eqnarray}
 {\cal L}_V &=&  \mu^2_{1}\Phi^{\dagger}_{1}\Phi_1+
\mu^2_{2}\Phi^{\dagger}_{2}\Phi_2+m^2_{12}\Phi^{\dagger}_{1}\Phi_2
+m^{*2}_{12}\Phi^{\dagger}_{2}\Phi_1+
\lambda_1(\Phi^{\dagger}_{1}\Phi_1)^2+ 
\lambda_2(\Phi^{\dagger}_{2}\Phi_2)^2 \nonumber \\
&&+\lambda_3(\Phi^{\dagger}_{1}\Phi_1)(\Phi^{\dagger}_{2}\Phi_2)
+\lambda_4(\Phi^{\dagger}_1\Phi_2)(\Phi^{\dagger}_2\Phi_1)
+\lambda_5(\Phi^{\dagger}_1\Phi_2)^2+\lambda^*_5(\Phi^{\dagger}_2\Phi_1)^2
 \nonumber \\
&&+\lambda_6(\Phi^{\dagger}_1\Phi_1)(\Phi^{\dagger}_1\Phi_2)
+\lambda^*_6(\Phi^{\dagger}_1\Phi_1)(\Phi^{\dagger}_2\Phi_1)
+\lambda_7(\Phi^{\dagger}_2\Phi_2)(\Phi^{\dagger}_1\Phi_2)
+\lambda^*_7(\Phi^{\dagger}_2\Phi_2)(\Phi^{\dagger}_2\Phi_1)
\end{eqnarray}
At tree--level $\lambda_1\to \lambda_4$ are functions of the
$SU(2)$ and $U(1)$ gauge couplings while
$\lambda_5=\lambda_6=\lambda_7=0$. In the 1-loop effective potential
all $\lambda_i$ receive sizeable corrections from the enhanced Yukawa
couplings of the third generation squarks. In particular 
$\lambda_5\ne \lambda_6\ne \lambda_7 \ne 0$ and are complex
in the presence of SUSY phases in $A_t$,$A_b$ and $\mu$. 
Explicit formulae may be found in \cite{NPB553}.

The restricted form of the tree--level potential limits the
possible mass splittings between $M_{H^\pm}$,$M_{H^0}$ and $M_A$.
It can be shown that $M_{H^\pm}$ and $M_A$ are related by:
\begin{equation}
M^2_{H^\pm}-M^2_{A}=(\lambda_4/2-Re(\lambda_5))v^2 \label{split}
\end{equation}
At tree--level   
$\lambda_4=g_W^2/2$ and $\lambda_5=0$, 
and the above relationship reduces to the familiar sum rule
\begin{equation}
M^2_{H^\pm}-M^2_{A}=M^2_W
\end{equation}
With the current LEP bound of $M_A\ge 90$ GeV one finds
$M_{H^\pm}-M_A\le 30$ GeV, with approximate degeneracy
for $M_A\ge 200$ GeV.

At the 1--loop level $\lambda_4$ and $\lambda_5$ may be written as
(keeping the dominant terms):
\begin{eqnarray}
\lambda_4 &\approx& {g_W^2\over 2}-{3\over 96\pi^2}\left[h_t^4\left(
{3|\mu|^2\over M^2_{SUSY}}-{3|\mu|^2|A_t|^2 \over M^4_{SUSY}}
\right)+h_b^4\left(
{3|\mu|^2\over M^2_{SUSY}}-{3|\mu|^2|A_b|^2 \over M^4_{SUSY}}
    \right)\right]          \nonumber\\ 
&&+{3\over {8\pi^2}}h_t^2h_b^2\left[{1\over 2}X_{tb}\right] \\ 
\lambda_5 &\approx&  {3\over{192\pi^2}}\left[
h_t^4\left({\mu^2 A_t^2\over M^4_{SUSY}}\right)
+ h_b^4\left({\mu^2 A_b^2\over M^4_{SUSY}}\right)\right] \nonumber
\label{eq:lam45}
\end{eqnarray}
where $X_{tb}$ is a function  of $A_t$,$A_b$, $M_{SUSY}$ and $\mu$.
From now on we will focus on the case of 
$\mu,A_{t,b}$ being real, with mention given to the case of complex
phases where appropriate. It can be seen from eq.~(4)
that $\mu,A_t$ or $\mu,A_b \ge 4M_{SUSY}$ would overcome much of the 
suppression from the small coefficients, permitting $\lambda_4,\lambda_5=
{\cal O}(1)$. From eq.~(2) this would give rise to
large mass differences $M_{H^\pm}-M_A$.

One can see from 
eq.~(2) that $M_{H^\pm} > M_A$ requires 
$\lambda_4 > \lambda_5$
while $M_{A} > M_{H^\pm}$ requires $\lambda_5 > \lambda_4$.
We note that both terms in $\lambda_5$ are positive definite
and the term proportional to $h_t^4$ will
dominate unless $\tan\beta$ is large (which enhances 
$h_b$). In the expression for 
$\lambda_4$ there may be both constructive and
destructive interference among the various terms  
and so the SUSY correction to the tree--level 
value may take either sign.

For the CP even Higgs bosons one has a mass
matrix ${\cal M}_S$ which when diagonalized gives the mass
eigenstates $h^0$ and $H^0$. Note that the 1--loop corrected
${\cal M}_S$ involves all $\lambda_i$, $i=1\to 7$, and 
its explicit form may be found
in \cite{NPB553}. Of interest to us is the
mass splitting between $M_{H^\pm}$ and $M_{H^0}$ which may
be given approximately by:
\begin{eqnarray}
  M^2_{H^\pm} - M^2_{H^0} \sim v^2 \left( 
       {\lambda_4 \over 2} +{\rm Re} \lambda_5 
   + { 2 {\rm Re } \lambda_6 \over \tan\beta} 
  \right).
\label{splita}
\end{eqnarray}

Recently it has been shown that the SUSY radiative corrections 
to the effective Yukawa couplings \cite{Yuk} can be 
important (for a review see \cite{logan}). Such 
corrections comprise loops involving gluino--sbottom and chargino--stop.
The modified $h_b$ has already
been shown to sizeably affect $b \to s \gamma$ 
\cite{giudice},\cite{nierste}, the effective $H^{\pm}tb$ coupling 
\cite{htb}, and the Higgs decay $h^0\to b\overline b$ \cite{Borzumati,loghab}.
In our analysis we shall restrict ourselves to $\tan\beta\le 20$, which
is the region where the above corrections to $h_b$ have minor impact.

Note that such large mass splittings occur more naturally
in extended versions of the MSSM which include a singlet 
Higgs field in the superpotential e.g. \cite{Panagiotakopoulos:2001zy}.
In such models eq.~(2) is modified
to include a term $\sim\lambda v^2$, where $\lambda=\mu/v_s$ ($v_s$ is
the vacuum expectation value of the singlet Higgs field). Thus
$\lambda={\cal O} (g_w)$ (i.e. gauge coupling strength)
may be attained with $\mu= {\cal O} (100)$ GeV,
giving rise to large mass splittings.

\section{Numerical results}
We now present our numerical results for the mass splittings
$M_{H^\pm}-M_{H^0}$ and $M_{A}-M_{H^\pm}$. 
We take two representative sets of parameters: 
set (A) gives large
$M_{H^\pm}-M_{H^0}$, and set (B) gives large $M_{A}-M_{H^\pm}$.
\begin{eqnarray}
\begin{array}{|c|c|c|c|c|c|c|c|c|}
\hline
   & \tan\beta & M_{H^\pm} & M_{\tilde{Q_3}} &  M_{\tilde{t}} 
& M_{\tilde{b}} & \mu & A_t & A_b   \\
\hline
\hline
(A) & 11  & 250 & 500 & 550 & 550 &  4000 & 1900 & 0   \\
\hline
(B) & 10  & 150 & 250 & 200 & 500 &  2800 & 0 & 0  \\
\hline
\end{array}
\label{sets}
\end{eqnarray}
Here all masses are in GeV;
$M_{\tilde{Q_3}},M_{\tilde{t}},M_{\tilde{b}}$ refer to
third generation squark soft SUSY breaking masses; $A_t$ and $A_b$
are the analogous trilinear couplings. We take $\mu$ and $A_{t,b}$
to be real in the numerical analysis.
For these values of $\tan\beta$ the terms
$\sim h^4_t$ dominate the terms $\sim h^4_b$ in the expressions for the
$\lambda_i$. We define $M_{SUSY}$ as
\begin{equation}
M^2_{SUSY}=(m^2_{\tilde t_1}+m^2_{\tilde t_2})/2
\end{equation}
where $\tilde t_1$ and $\tilde t_2$ refer to the lighter and heavier
stop eigenstates respectively. For set(A) $M_{SUSY} \approx 450$ GeV 
and for set(B) $M_{SUSY} \approx 280$ GeV. For both parameter sets 
the mass splitting  
$m^2_{\tilde t_2}-m^2_{\tilde t_1}$ comfortably satisfies the 
RG analysis requirement \cite{NPB553}
\begin{equation}
{m^2_{\tilde t_2}-m^2_{\tilde t_1}\over 
m^2_{\tilde t_2}+m^2_{\tilde t_1}} \lesssim 0.5
\end{equation}
Note that the large value of $\mu$ taken in set (A) and (B)
does not imply an
unacceptably light $\tilde t_1$, since the corresponding 
entry in the stop mass matrix is $\mu\cot\beta$, which is comfortably
suppressed for the assumed values of $\tan\beta$. Larger
values of $\tan\beta$ together with large $\mu$ would invariably
generate values of $m_{\tilde b_1}$ which are lighter than the current
experimental bounds. We also require that all $\lambda_i$ remain
in the perturbative region.

In Fig.~1 (a) and (b), we plot $M_{H^\pm}-M_{H^0}$ and
$M_{A}-M_{H^\pm}$ as a function of $\tan\beta$
for $\mu$=$4500,4000,3500$ GeV (from above)
and $\mu$=$3300,2800,2300$ GeV (from above), respectively.
For the other parameters we take the choices in (\ref{sets}).

In both Figs.~1 and 2, the line in thin typeset 
signifies that the corresponding mass for one of 
$\tilde t_1$, $\tilde b_1$ and $M_{h^0}$
becomes smaller than the present experimental bound.

In Fig.~1(a), we can understand the behaviour of the
splittings $M_{H^\pm}-M_{H^0}$ from eqns.(4) and (\ref{splita}).
In eq.~(4), increasing $\tan\beta$ enhances the 
term $\sim h_b^2h_t^2$ which is negative in this parameter space, 
and thus reduces the positive $\la_4/2$. $\la_5$ is also positive 
and is not so
sensitive to $\tan\beta$ since the top quark Yukawa term dominates.
However, $2 \la_6/\tan\beta$ is negative and its modulus 
becomes smaller as $\tan\beta$ increases. The combined effect is
as follows:
for small $\tan\beta$ the negative term in eq.~(5)
cancels the positive terms and the splitting is
small. As $\tan\beta$ increases the positive terms dominate, giving
rise to large splittings. Further increases in $\tan\beta$ reduce the
still positive $\lambda_4$, while $\lambda_5$ remains approximately
constant, which explains the descent of the curve for larger $\tan\beta$. 
We have also checked that the splittings are relatively insensitive
to variations in $A_b$ since the relevant terms are suppressed by
$h_b^4$. One might expect that the splittings could be further increased by 
changing the
sign of $A_t$, which makes $\la_6$ positive. However it is not easy
to satisfy the scalar quark mass bound with this choice
in the CP conserving scenario. Larger values of $\mu$ give rise to larger
splittings, which is explained in Fig.~2(a) below.

In Fig.~1(b) one can see that $M_{A}-M_{H^\pm}$ increases with $\tan\beta$, 
which is explained by the fact that $\lambda_4$ is decreasing (see above)
and thus enhances the splitting from eq.~(2). Large $\mu$ provides the
largest splittings (see below).

In Fig.~2 (a) we plot $M_{H^\pm}-M_{H^0}$
as a function of $\mu/M_{SUSY}$
for $A_t =2000, 1900, 1800$ GeV (from above). 
For the other parameters we take the choices in (\ref{sets}).
In Fig.~2 (a), we can see that $\mu/M_{SUSY} \gtrsim 6$ allows
splittings $M_{H^\pm}-M_{H^0} \gtrsim 80$ GeV. Increasing $A_t$ 
enhances the positive contributions to 
$\la_4$ and $\la_5$, which in turn enhances $M_{H^\pm}-M_{H^0}$ 
(from eq. (\ref{splita})).

In Fig.~2 (b), we plot $M_{A}-M_{H^\pm}$ as a function of $\mu/M_{SUSY}$
for $M_{H^\pm} =150, 250, 350$ GeV (from above), since it turns out that
the splittings are not very sensitive to $A_t$.
As expected, we can see the decoupling behaviour, with the splitting
decreasing for increasing $M_{H^\pm}$.
For low values of $\mu/M_{SUSY}$
the 1-loop corrections are not so large and so one has approximately
the tree--level result $M_{H^\pm}\ge M_A$. As $\mu/M_{SUSY}$ increases,
$\lambda_5$ is enhanced
to large positive values. In addition, $\lambda_4$ is
decreased from its positive tree-level value.
Hence the mass splitting $M_A-M_{H^\pm}$ increases and exceeds
80 GeV for $\mu/M_{SUSY}\gtrsim 10$.

If CP violating phases are allowed in
$\mu$ and $A_{t,b}$ then $\la_{5,6}$ become complex numbers in general.
The mass eigenstates $H_2^0$ and $H_3^0$ are now mixed states
of CP, and the large splittings $M_{H^\pm}-M_{H_2^0}$ or
$M_{H_3^0}-M_{H^\pm}$ are possible in a wider range of parameter
space which satisfies the mass bounds on $H_1^0$ and $\widetilde{t}_1$
($\widetilde{b}_1$). 

The magnitude of the mass splittings has important consequences
for the phenomenology of the MSSM. Our results show that
$M_{H^\pm}$ need not be close in mass to $H^0$ and $A^0$, and the
mass differences $|M_{H^\pm}-M_{H^0,A}|$ may be as large as the
analogous values in a general 2HDM or in other extended Higgs sectors. 
The relation $M_{H^\pm}\approx M_{H^0,A}$ for $M_{H^\pm}\ge 200$ GeV, 
which is assumed in existing phenomenological analyses,
may be broken in the region of large $\mu/M_{SUSY}$. 
We stress that the large values of $\mu$ considered here do not arise
in popular SUSY models, such as
minimal supergravity or gauge mediated models. As discussed in
\cite{plb495}, in order to be compatible
with electroweak symmetry breaking, such large values of $\mu$ would require 
the soft SUSY breaking mass parameters $m^2_1$ and $m^2_2$ to be
of the order of $|\mu|$ and negative, and considerable fine-tuning is
necessary.

If $M_{H^\pm}-M_{H^0}\ge 80$ GeV then the 2 body decays
$H^{\pm}\to H^0 W^\pm$  would be open. This would offer a 
new discovery channel for $H^\pm$ at the LHC, and one which is 
expected to offer a very promising signature. Ref.~\cite{HAW} presented
a signal--background analysis showing that the background is small, and
any model which allows a large BR($H^{\pm}\to H^0W^\pm$) would
provide a very clear signal in this channel. 
We shall show below that such large
BRs are possible in the MSSM, and consequently would aid the search 
for $H^\pm$ at the LHC.
  
If $M_{A}-M_{H^\pm}\ge 80$ GeV then the 2 body decays
$A^0\to H^\pm W^\mp$ would be open. This would have important
consequences for the process 
$\mu^+\mu^-\to H^\pm W^\mp$ at a muon collider, which receives
contributions from $A^0,H^0$ mediated s--channel diagrams, and  
was shown to be a promising production mechanism for $H^{\pm}$
in the MSSM with $M_{H^\pm}\approx M_{H^0,A^0}$ \cite{AAD}. 
Any splittings with $M_{A}\ge M_{H^\pm}$ 
would enhance the rate for $\sigma(\mu^+\mu^- \to H^{\pm}W^{\mp}$)
compared to that in \cite{AAD}. For $M_{A}-M_{H^\pm}\ge 80$ GeV
one could have resonant $H^\pm$ production via
$\mu^+\mu^-\to A^0\to H^\pm W^\mp$, which was shown to allow
very large cross--sections ($>1$ pb) in the case of the general 
2HDM \cite{mumuHW}. These possibilities will be pursued in a 
future article \cite{progress}.

In Fig.~3 we plot branching ratios (BRs) for the decay
$H^{\pm}\to H^0 W^\pm$ as a function of $\tan\beta$.
This two body decay is open if $M_{H^\pm}-M_{H^0}\ge 80$ GeV,
and is proportional to $\sin^2(\beta-\alpha)$. Also plotted
is BR($H^{\pm}\to h^0W^\pm$) which $\sim \cos^2(\beta-\alpha)$. 
We take the same parameters as Fig.~1 (a). 
The thick (thin) lines represent BR($H^{\pm}\to H^0 (h^0) W^\pm$)
for $\mu=4500, 4000, 3500$ (from above), respectively.
Note that we do not consider decays to SUSY particles 
$H^{\pm}\to \tilde t \tilde b$, $\chi^\pm\chi^0$ etc. The decays
$H^\pm\to \tilde t_1\tilde b_1$ are not open for this choice of
$M_{H^\pm}$ and $H^\pm \to \chi^\pm\chi^0$ can be 
closed by choosing suitable values for $M_1$,$M_2$ etc.
It can be seen from Fig.3 that BR$\gtrsim 20\%$ is 
attainable for intermediate values of $\tan\beta$, clearly
showing the importance of this new decay channel. 

The behaviour of the BRs can be qualitatively understood as follows:
for low $\tan\beta$ the splittings are not large enough to open
the channel and for large $\tan\beta$ the newly opened channel
$H^{\pm}\to H^0 W^\pm$ cannot compete with the fermionic decay modes.
Thus the maximum inpact of $H^{\pm}\to H^0W^\pm$ decays
is for intermediate values of $\tan\beta$, which is the
most problematic region for the $H^{\pm}$ search at the LHC.
The current search strategies
utilize $H^{\pm}\to \tau\nu_{\tau}$ decays \cite{Roy}, which is most 
effective for $\tan\beta\ge 15$, or $H^\pm\to tb$ decays which cover
the regions $\tan\beta\le 3$ and $\tan\beta\ge 25$ \cite{Atlas9913}.
We can also see that BR($H^{\pm}\to h^0 W^\pm$), 
although suppressed in the decoupling limit, can
be significant. The sizeable BRs shown in Fig.~3 would offer 
very good detection prospects in these channels \cite{HAW}.

In general, the mass splittings presented here may give rise
to a large contribution to the precisely measured electroweak 
parameter $\rho$. For mass
splittings $\ge 20$ GeV we find that the Higgs contribution is always
positive and violates the $\delta\rho$ constraint at the $2(3)\sigma$
level for $|M_{H^\pm}-M_{H^0,A}|\ge 150(200)$ GeV.
The dominant SUSY particle contribution ($\tilde t-\tilde b$ loops)
to $\delta\rho$ has the same sign 
as the Higgs contribution \cite{hagiwara}, 
thus further reducing the maximum allowed value of
$|M_{H^\pm}-M_{H^0,A}|$. Therefore the mass splittings
$|M_{H^\pm}-M_{H^0,A}|$ presented here are
consistent with the $\delta\rho$ constraints.

\section{Conclusions}
We have shown that very large mass splittings  $M_{H^\pm}-M_{H^0,A}$ 
are possible in the MSSM. Such splittings occur in a previously ignored
region of the MSSM parameter space and may exceed 100 GeV,
thus strongly violating the commonly assumed degeneracy relation
$M_{H^\pm}\approx M_{H^0}\approx M_{A}$. 
The largest splittings arise for relatively large 
$\mu\ge 6M_{SUSY}$. The previously neglected
2 body decays $H^{\pm}\to H^0W^\pm$ and $A^0\to H^+W^-,H^-W^+$
may become the dominant channels, which would have important
consequences for Higgs searches at future colliders.  
If $M_{A}-M_{H^\pm} \gtrsim 80$ GeV, a muon collider 
could copiously produce $H^\pm$ at resonance.  

\section*{Acknowledgements}
We thank A. Arhrib, P. Ko and Y. Okada for useful comments. We
are grateful to C. Wagner for confirming some of the results 
presented here.


\newpage
\begin{figure}[t]
\smallskip\smallskip  
\centerline{{\hskip0.352cm\epsfxsize3 in \epsffile{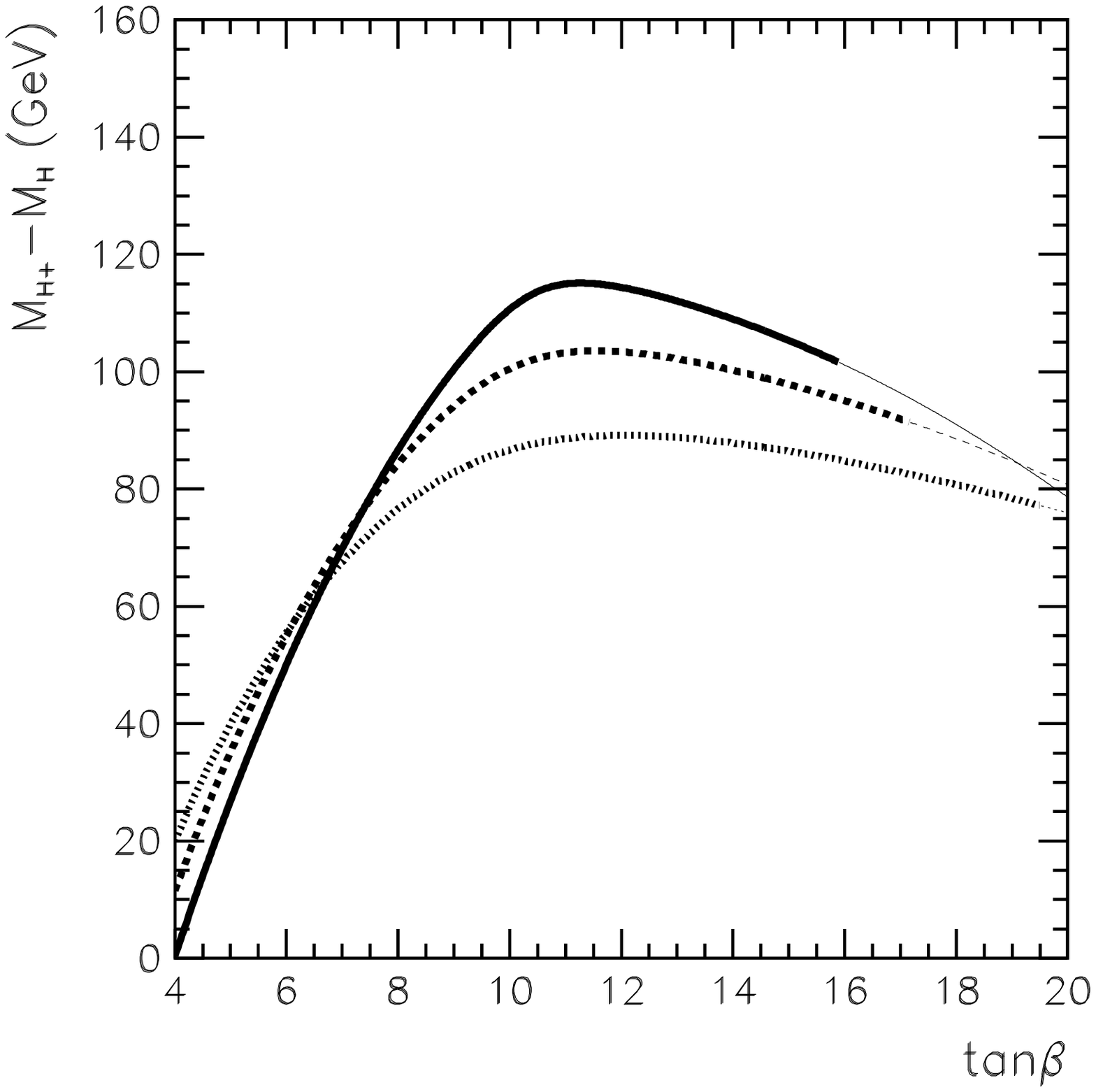}}
\hskip0.4cm{\hskip0.352cm\epsfxsize3 in \epsffile{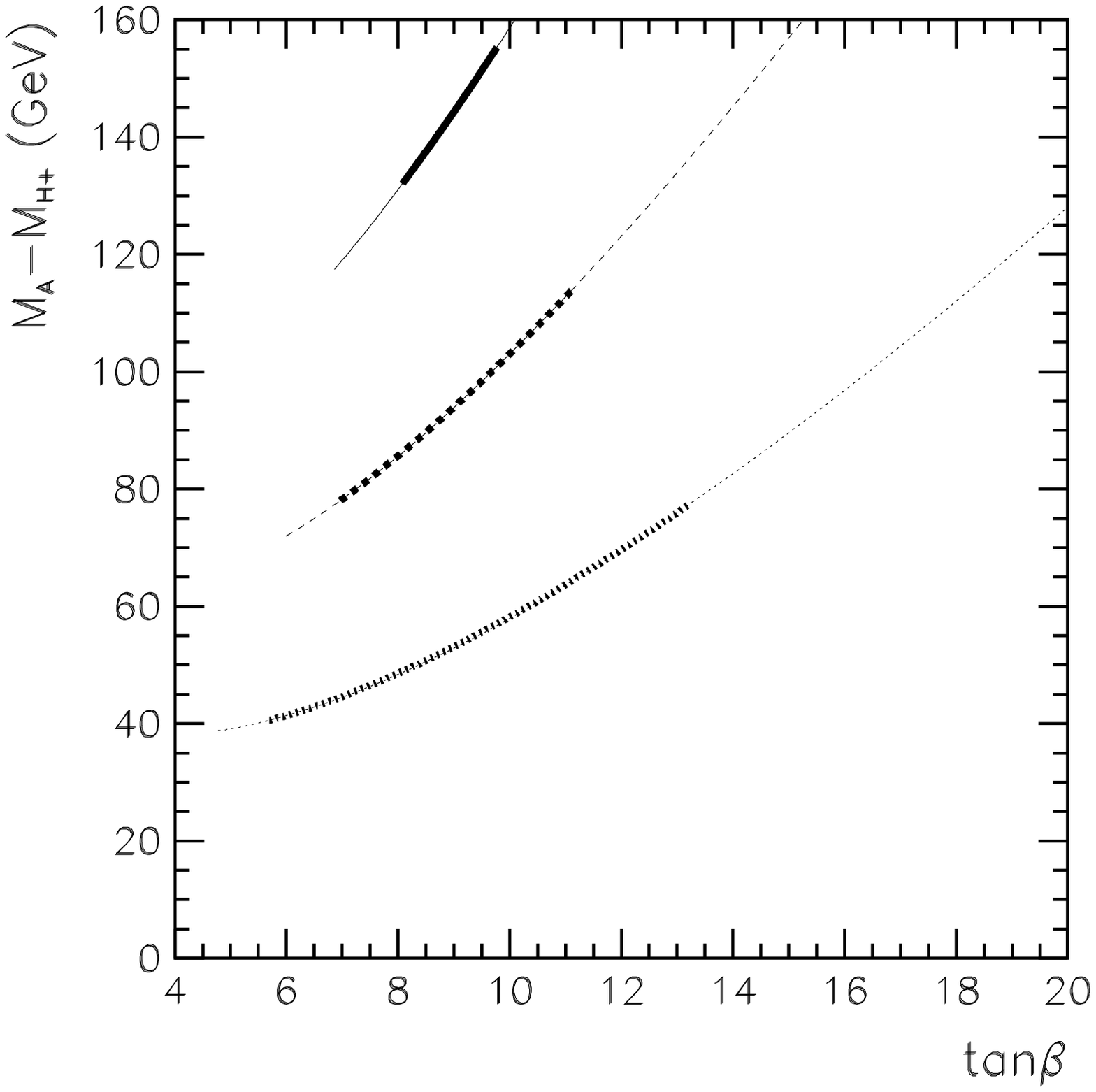}} }
\smallskip\smallskip\smallskip\smallskip
\caption{
$M_{H^\pm}-M_{H^0}$ (left panel) and
$M_{A}-M_{H^\pm}$ (right panel) as a function of $\tan\beta$
for $\mu=4500, 4000, 3500$ GeV (from above, left panel)
and $\mu=3300, 2800, 2300$ GeV (from above, right panel), respectively.
The thin lines violate the experimental mass bounds on
$h^0$, $\widetilde{t}_1$ or $\widetilde{b}_1$.
For other parameters, see the text.
}
\label{fig1}
\end{figure}

\begin{figure}[b]
\smallskip\smallskip  
\centerline{{\hskip0.352cm\epsfxsize3 in \epsffile{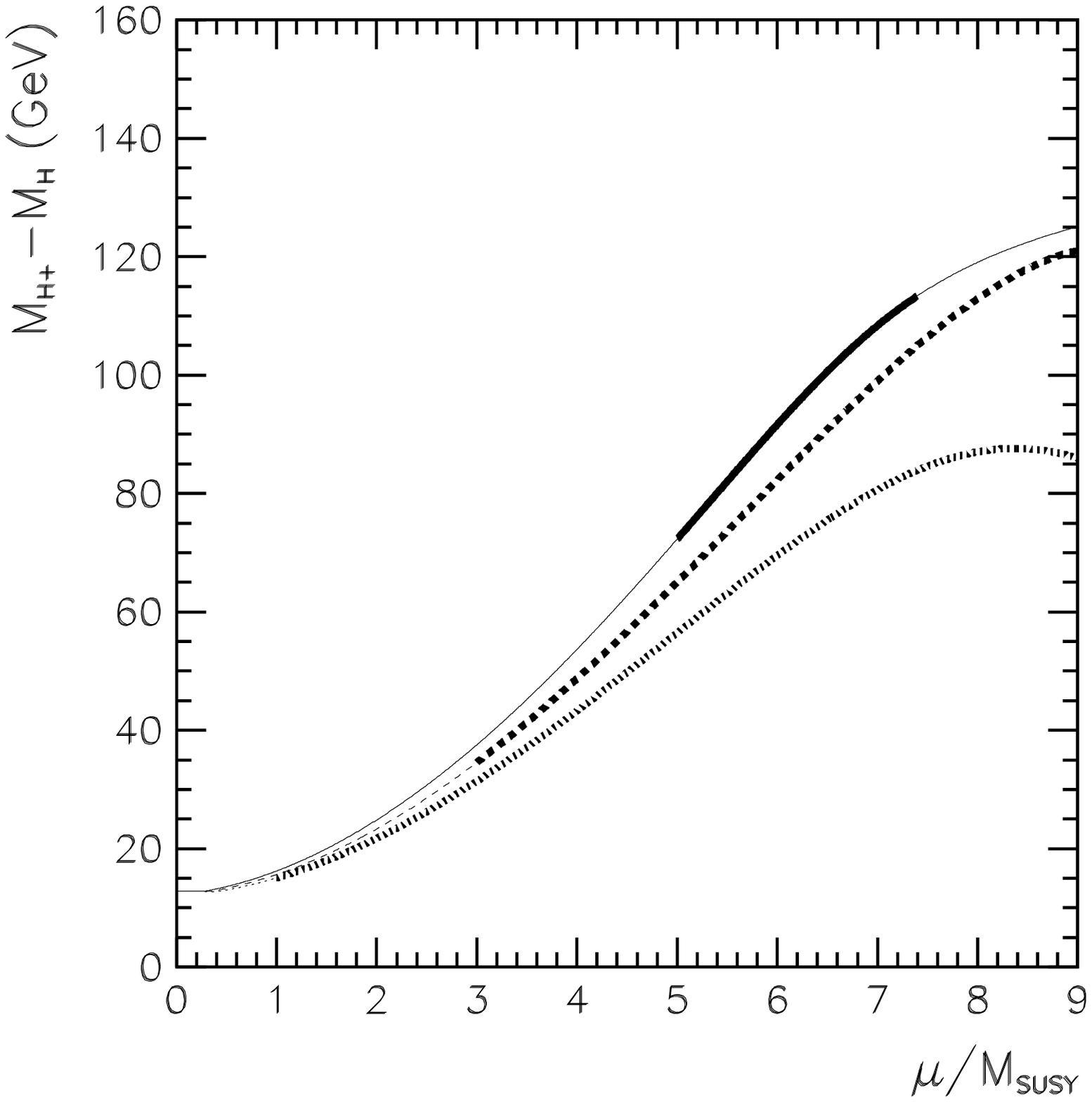}}
\hskip0.4cm{\hskip0.352cm\epsfxsize3 in \epsffile{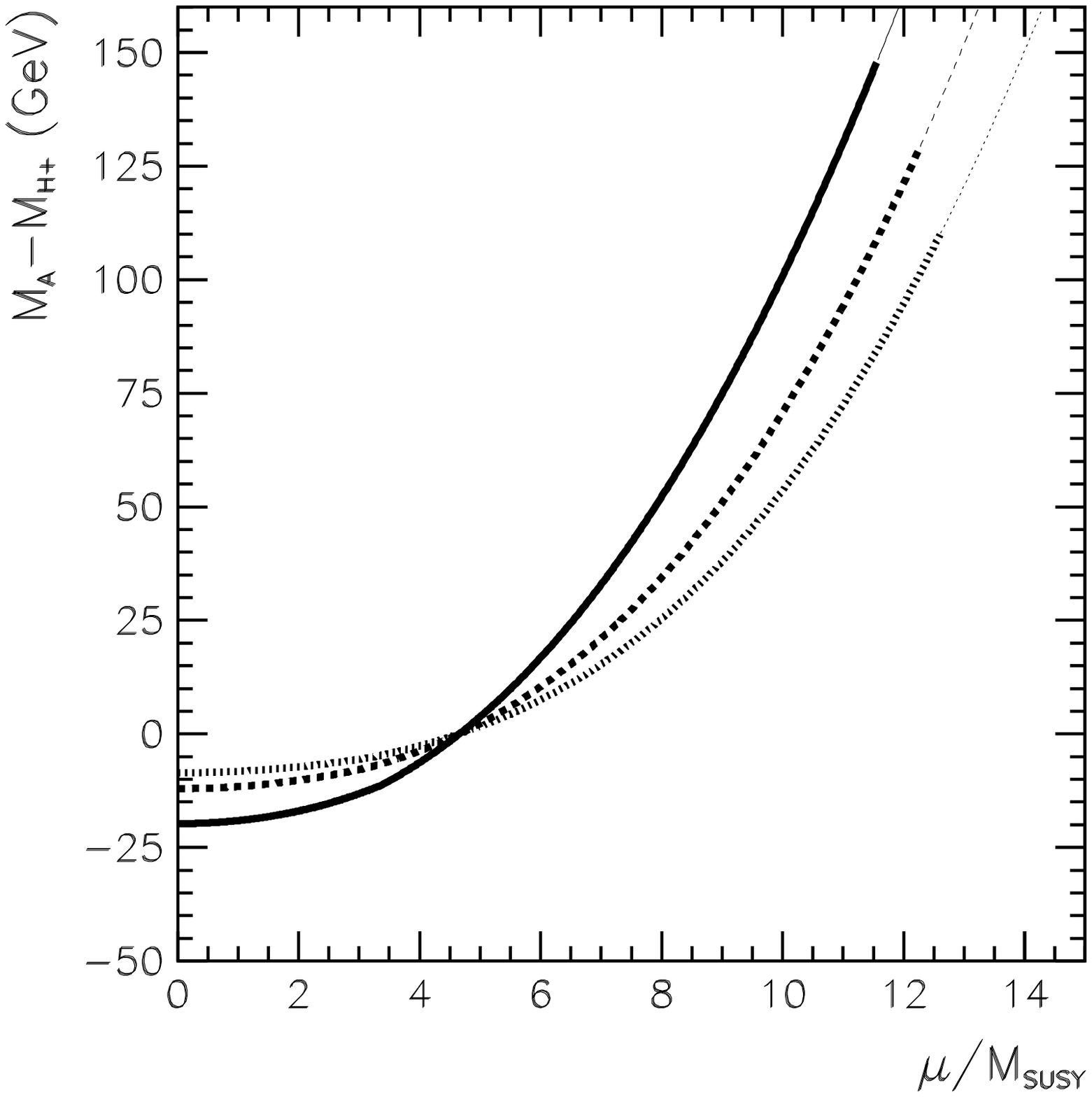}} }
\smallskip\smallskip\smallskip\smallskip
\caption{
$M_{H^\pm}-M_{H^0}$ (left panel) and
$M_{A}-M_{H^\pm}$ (right panel) as a function of $\mu/M_{SUSY}$
for $A_t=2000, 1900, 1800$ GeV (from above, left panel)
and $M_{H^\pm} = 150, 250, 350$ GeV (from above, right panel), respectively.
The thin lines violate the experimental mass bounds on
$h^0$, $\widetilde{t}_1$ or $\widetilde{b}_1$.
For other parameters, see the text.
}
\label{fig2}
\end{figure}

\begin{figure}[t]
\smallskip\smallskip  
\centerline{{\hskip0.352cm\epsfxsize4 in \epsffile{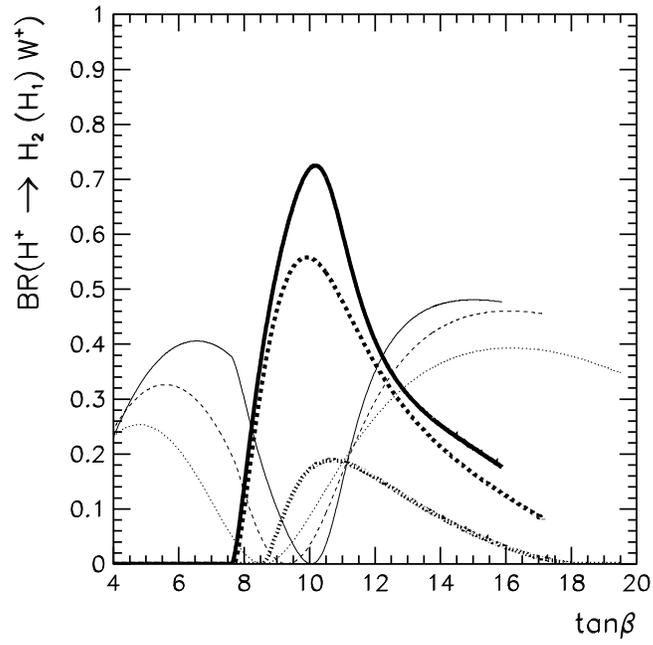}}
}
\smallskip\smallskip\smallskip\smallskip
\caption{BR($H^{\pm}\to H^0(h^0) W^\pm$) (thick (thin) lines) as a function
of $\tan\beta$ for the same parameter choice as Fig.~1 (a).
}
\label{fig3}
\end{figure}

\end{document}